# DeepBreath: Deep Learning of Breathing Patterns for Automatic Stress Recognition using Low-Cost Thermal Imaging in Unconstrained Settings

Youngjun Cho, Nadia Bianchi-Berthouze
*UCL Interaction Centre, Faculty of Brain Sciences*
*University College London*
*London, United Kingdom*
{youngjun.cho.15, nadia.berthouze}@ucl.ac.uk

Simon J. Julier
*Department of Computer Science*
*University College London*
*London, United Kingdom*
s.julier@ucl.ac.uk

*Abstract*—We propose DeepBreath, a deep learning model which automatically recognises people's psychological stress level (mental overload) from their breathing patterns. Using a low cost thermal camera, we track a person's breathing patterns as temperature changes around his/her nostril. The paper's technical contribution is threefold. First of all, instead of creating hand-crafted features to capture aspects of the breathing patterns, we transform the uni-dimensional breathing signals into two dimensional respiration variability spectrogram (RVS) sequences. The spectrograms easily capture the complexity of the breathing dynamics. Second, a spatial pattern analysis based on a deep *Convolutional Neural Network (CNN)* is directly applied to the spectrogram sequences without the need of hand-crafting features. Finally, a data augmentation technique, inspired from solutions for over-fitting problems in deep learning, is applied to allow the CNN to learn with a small-scale dataset from short-term measurements (e.g., up to a few hours). The model is trained and tested with data collected from people exposed to two types of cognitive tasks (Stroop Colour Word Test, Mental Computation test) with sessions of different difficulty levels. Using normalised self-report as ground truth, the CNN reaches 84.59% accuracy in discriminating between two levels of stress and 56.52% in discriminating between three levels. In addition, the CNN outperformed powerful shallow learning methods based on a single layer neural network. Finally, the dataset of labelled thermal images will be open to the community.

## 1. Introduction

Our modern society suffers from psychological stress. Although the literature provides several definitions of stress, it is evident that various forms of stress influence our mental and physical health [1]–[3]. In this paper, we aim to contribute to the body of work on automatic detection of people's stress levels for the long-term goal of technology-based stress management interventions (e.g., [4], [5], [6]). We propose *DeepBreath*, a deep learning model to automatically infer stress levels (mental overload in our study) from thermal imaging of people's breathing patterns in unconstrained settings. Our novel automatic stress recognition system makes three contributions: 1) a novel representation of breathing pattern dynamics; 2) a deep learning framework which does not require feature hand-crafting and a data augmentation technique to learn with a small-scale dataset; 3) a proposal of low-cost thermal imaging as a ubiquitous stress sensor. These contributions are motivated below.

Given that mental stress, defined here as *"a load producing a complex pattern of effects on a body and a mind"* [7], affects physiological processes [2], [8], [9], breathing among other physiological processes have been explored for automatic detection of stress (e.g., [6], [10], [11]). However, most of these works focuses on gross statistical features. In particular, in the case of breathing, average breathing rate (over a specified window) is the main features used [10], [11]. However, according to Grossman [9], breathing irregularity can also be observed in stressful situations. Here, we propose the use of bi-dimensional spectrogram variability sequences to capture in a compact way the variability of breathing patterns over time.

Our second contribution relies on the use of deep learning to eliminate the need for feature handcrafting. As discussed in [12], even carefully hand engineered-feature extractors could fail to generalize to unseen data sets. Hence we have turned to deep learning [12] to automatically find good features during the learning process. In this paper, we propose a novel approach to the *"deep learning"* of the relationship between the variability spectrogram sequences and mental stress levels. A key problem with deep learning is the requirement for the considerable amount of training data, which can be difficult and time consuming to obtain. To overcome this problem, we finally propose to use a unidirectional sliding cropper-based data augmentation algorithm inspired by deep learning solutions for over-fitting.

Finally, our third contribution lays in the type of sensor used. Typically used technology for breathing tracking requires to wear sensors (e.g., chest belt, bracelets) [11] or to use RGB cameras [10] that raise privacy issues as well as lighting challenges. Thermal imaging can be free from those constraints while supporting breathing measurements [13]–[15]. Despite our approach to automatic stress detection is sensor-independent, in this paper we investigate it through the use of a low-cost mobile thermal camera which was first investigated in [15] and is still quite an underexplored sensor in affective computing.

The rest of the paper is organized as follow: we review the background work and present the technical details of the proposed DeepBreath method. Next, we summarise our experimental protocol and results. We conclude by discussing and comparing the performances of our new approach against shallow learning.

## 2. Background

### 2.1. Breathing and Stress

Breathing is an important vital process controlled by the Autonomic Nervous System (ANS). The monitoring of breathing patterns can be informative of a person's mental and physical condition. Researchers have investigated the possibility of using breathing signals together with other physiological signatures to automatically assess people's stress level [6], [10], [11]. However, the results of these studies are unclear about the contribution of breathing features as compared to other physiological measures. For instance, with 85% accuracy, McDuff *et al.* [10] show that in the binary classification of stress (i.e., rest and stress) using multiple physiological features (i.e., heart rate, breathing rate and heart rate variability (HRV)) the most informative feature is HRV with breathing rate playing a minor role. It is possible that breathing rate itself is not the best discriminating feature for respiration. Indeed, the direction of changes in respiration rate during stress is not so clear. For instance, while Masaoka *et al.* [16] identified an increase of breathing rate during mental stress, Hong et al. [11] reported a severe drop of the rate along with the stress. In addition, when exploring HRV, multiple parameters are computed to capture its complex dynamics. Given these results, we propose to explore the complexity of breathing patterns rather than just discrete rate.

### 2.2. Thermal imaging and affect detection

Thermal imaging is a key non-contact method to study heat patterns of materials and organisms. Despite thermographic channels being still little explored in affective computing, various studies have explored the possible thermal signatures appearing in association with a person's psychological affective states (e.g., [17]–[21]). For instance, Engert *et al.* [18] identified that a decrease of temperature of the nose tip and perioral areas could be a barometer of mentally stressful states. While most works (e.g., [17]–[19]) focused on attempting to confirm the relationship between directional changes in temperature of facial areas and affective states, a few works investigated how such thermal information can be used for automatic affect recognition (e.g., [20], [21]). For example, Nhan and Chau [20] used temperature changes on left, right supraorbital, periorbital, and nasal areas as features for distinguishing high arousal from a baseline.

All the earlier works, however, employed very heavy and expensive thermographic systems, not easily to set up in any position and not easily portable. In addition, tools for thermal image processing (e.g., automated tracking of a Region-Of-Interest (ROI)) were also limited (e.g., a dot stick was used for the tracking of a ROI in [20]), constraining its deployments (e.g., no head movement allowed, environments with stable temperature only). However, in a recent study [15], we have proposed a new ROI tracking method based on the Optimal Quantization technique and the Thermal Gradient Flow algorithm reaching extreme robustness in automatic tracking of breathing patterns in unconstrained settings comparable to the standard breathing belt. In addition, our study shows that the algorithm works with low-cost thermal cameras (e.g., Seek Thermal, FLIR One), portable and hence more suitable for real-life situations. We propose to build on this method to test the possibility of using low-cost camera to automatically detect stress levels from breathing during unconstrained sedentary mental tasks.

### 2.3. Deep learning for affective computing

*Deep learning*, which is currently one of the most successfully and widely used approaches in Computer Vision and Pattern Recognition (CVPR) [12], [22], [23], could be a potential solution for better understandings of physiological and behavioural patterns in relation to a person's affective states (e.g., distress, anxiety). Earlier deep learning models have been explored to detect facial and voice affective expressions. For instance, Liu *et al.* [24] boosted the deep belief network for better characterization of features in relation to a certain type of facial expressions. Lane *et al.* [25] proposed a model which could analyse multiple large scale speech datasets using a coupled deep neural network for automated stress detection. These approaches have achieved greater performances in the classification problems, in comparison with shallow learning approaches (e.g., SVM, Gaussian mixture models).

More recently, researchers have shown the power of Convolutional Neural Network (CNN) based models (e.g., [22], [23], [26], [27]) in a variety of problems spanning from image classification to semantic segmentation, demonstrating that the models themselves can produce better learning performances than earlier deep learning approaches. These recent accomplishments in CNNs have been adopted to improve affect recognition performances for large scale datasets: for instance, arousal and valence prediction in audio [28], facial action unit identification [29], and facial expression recognition in video [30]. However, the small size of datasets generally available in affective computing, mostly involving the uni-dimensionality of physiological signals, do not easily allow the use of such techniques. Thus, in this paper, we propose a new two-dimensional representation of breathing patterns capable of capturing the complexity of the patterns without feature crafting, and a data augmentation algorithm to enable the use of the CNN.

## 3. DeepBreath: Design and Algorithm

This section presents the process used to collect, represent and augment the data, and describes the deep learning architecture.

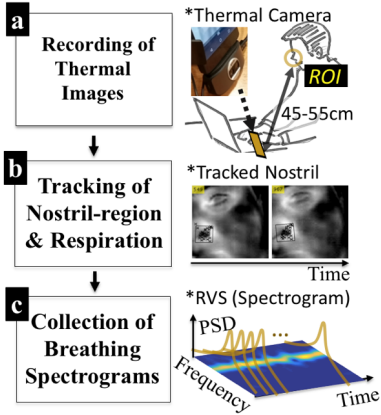

Figure. 1. Process for the respiration variability spectrogram collection through thermal imaging.

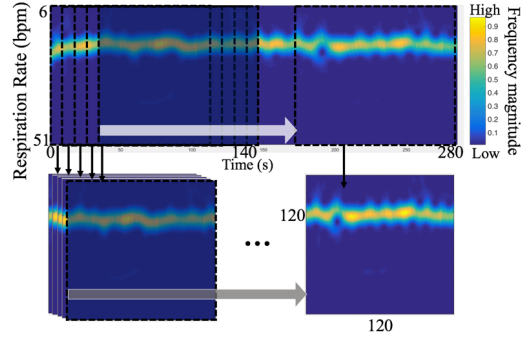

Figure. 2. Proposed data augmentation technique using a uni-directional sliding cropper with a 120x120 square window. The colour scale indicates the frequency magnitude (i.e., PSD).

### 3.1. Respiration Variability Spectrogram (RVS) collection from thermal imaging

Figure 1 illustrates the process for collecting breathing patterns using a portable low-cost thermal imaging camera (dimension: 18mm x 26mm x 72mm). Thermal videos are captured using the camera attached to a smartphone placed at a maximum distance of 55cm from a person's nostril area (i.e., our ROI) (figure 1a). We use the respiration tracking algorithm proposed in [15] to recover one-dimensional breathing signals from the nostril ROI on the thermal videos (see figure 1b).

Figure 1c shows a two-dimensional spectrogram automatically transformed from one-dimensional signal sequences. This is for condensing dynamic information (i.e., respiration variability) into an image which we call the RVS. More specifically, a two dimensional spectrogram can be constructed by stacking a Power Spectral Density (PSD) vector of a short-time-window respiration signals (i.e., Equation (11) in [15]) over time until the end of recordings. The PSD function handles the short-time autocorrelation that identifies similarities between neighbouring signal patterns, being of use to examine respiration variations in a short period. The RVS can be expressed as:

$$RVS(f,\hat{t}) = \sum_k R_{ww}(k,\hat{t})e^{-j2\pi fk} \quad (1)$$

where $R_{ww}(k,\hat{t})$ is the short-time autocorrelation output of windowed breathing signals from Equation (10) in [15], $k$ is a time lag to examine the similarity, $\hat{t}$ is the discrete time information (e.g., $\hat{t} \in [1s,100s]$), and $f$ is the frequency domain information.

For the implementation, as used in [15], the normal breathing frequency range of healthy adults, [$f_1$=0.1Hz, $f_2$=0.85Hz], was used to band-pass filter the input breathing signals in (1) for reducing magnitudes of frequencies beyond the range of interest. $f$ in (1) has the same range scale. The time length of the short-time window is set to 20 seconds and moves forwards in 1 second intervals. Lastly, so as to produce an integer RVS matrix (i.e., to be easily taken as a grey-image), we converted the frequency in [$f_1, f_2$] into an integer in a new scale [$1, m$] (here, $m$=120) using a linear transformation $y = T(f)$. Note that this new bi-dimensional input signature can be applicable to any types of breathing measurements (e.g., chest-belt, expensive high-definition thermal camera), in spite of our focus laid on low-cost low-definition thermal imaging.

### 3.2. Data augmentation

In comparison with other physiological signatures (e.g., heart rate, EEG), breathing patterns usually have a narrow and low frequency bandwidth (e.g., between 0.1Hz and 0.85Hz). In other words, it requires the collection of longer data sets to allow for a deep learning process to take place. One possibility is to use a pre-trained network which is trained on a larger dataset, which is known as an efficient way to learn small datasets [12]. However, low-level features learnt from other domains (e.g., ImageNet data [22]) may not fit breathing dynamic patterns. Inspired by basic transformation-based data augmentation methods in deep learning (e.g., cropping, zoom-in) [22], we propose a unidirectional sliding cropper with a square window to augment the RVS dataset while preserving each label, defined as:

$$I_i(x,y) = RVS(T^{-1}(y), i+x), \ x \in \{1,2,...,m\} \quad (2)$$

where $i \in \{1,2,...,\hat{t}_{max}-m+1\}$ is in an ascending order (1 second interval), $m$ is the length of each unit image, $y$ in (2) is inversely transformed (T$^{-1}$) to $f$ (in (1)), in turn producing $(\hat{t}_{max}-m+1)$ images of size $m$ x $m$ x 1 (height x width x channel). With $m$ set to 120 in Section 3.1, each unit spectrogram image has 120x120 size (Figure 2).

### 3.3. CNN architecture

The architecture of our network is shown in Figure 3. It consists of two convolutional layers, two pooling layers and one fully connected layer. Although additional modifications can be applied to this basic form of CNN [27], our aim in this paper is to investigate the possibility of using deep learning to classify stress levels from physiological patterns. In addition, we aim to build a low computational cost deep learning network for running on an embedded or mobile system (e.g., Figure 5a). Each image patch of 120 x 120 is resized to 28 x 28 using a basic bi-cubic interpolation to be fed forward to the first convolutional layer. The size of the resized patch was chosen to use the CNN structure proposed in [27]: The first layer filters the input image with $n$ kernels (this value is set along with the number of stress levels for classification in Section 5) of size 5 x 5, the second convolutional layer has $j$ kernels of size 5 x 5 and each pooling layer has averaging filters of size 2 x 2 applied with a stride of 2. A sigmoid function, widely used in artificial neural networks, is connected to every convolutional and fully connected layer to induce non-linearity. In the case where a large dataset is available, a Rectified Linear Unit (ReLU) can replace this to boost the learning speed [22]. The output of the final pooling layer is fully connected to each output neuron which corresponds to each targeted class. Stochastic gradient descent (SGD) is applied to supervised fine-tuning during back propagation to reflect classification errors on training data sets.

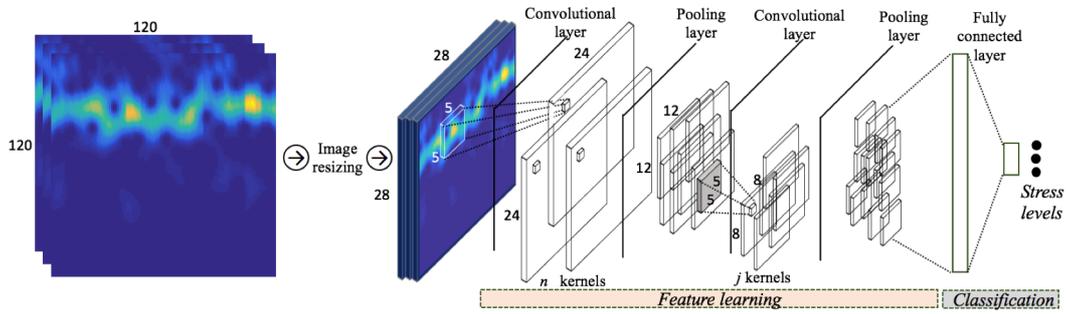

Figure. 3. The proposed CNN architecture consisting of two convolutional layers, two pooling layers and one fully connected layer.

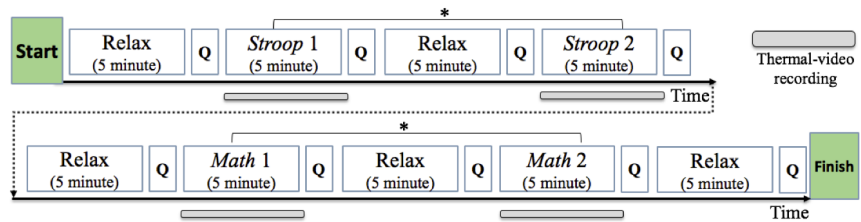

Figure. 4. Flow chart of the conducted study (*Counterbalanced in Latin squared design between easy and diffcult sessions).

## 4. Study Design and Dataset

To test the proposed approach, we have designed an experiment to induce different levels of stress in participants. The participants' breathing patterns were continuously recorded during the experiment and in turn constitute the dataset (participants' nostril area) for our study. The study procedure was approved by the Ethics Committee of University College London Interaction Centre.

### 4.1. Stress induction tasks

Two widely used tasks for inducing mental stress were selected for our purpose: the Stroop Colour Word test [31] (e.g., used in [32]) (denoted as *Stroop*) and the mathematics test (e.g., used in [10], [11]]) (denoted as *Math*). Each task has both an easy and a difficult session to ensure a good spread of induced stress levels within each task. This was important as the tasks differed in the amount of verbal output and behaviour they required. In the Stroop task, all participants had to name the colour of a word. In the easy session of the first task, the meaning of a word and its font colour were all congruent (e.g., the word *red* written in *red*). In the difficult session, these were incongruent (e.g., the word *yellow*, but written in *red*). The Math task required the participants to repeatedly subtract (mentally) a certain number (e.g., 1, 13) from a four digit number (e.g., 5000): while the subtracted number was set to a two digit number (e.g., 13) in the difficult session of the second task, that was set to 1, transforming the subtraction test to an easy counting-down test for the easy session. It was expected that the difficult sessions in Math and Stroop tests would lead to higher cognitive load than the easy sessions. After each answer, participants received sound feedback to inform them whether the answer was correct or not. Before and after each of the task sessions, the participants were asked to fill a short questionnaire (denoted as Q in Figure 4) to report their stress level. All the tests were programmed and run in MATLAB (2015b, The MathWorks). The program sources are publicly released at *http://youngjuncho.com/2017/ACII2017-open-sources/*.

### 4.2. Participants and procedure

8 healthy adults (3 females) (aged 18-53 years, M=30.75,

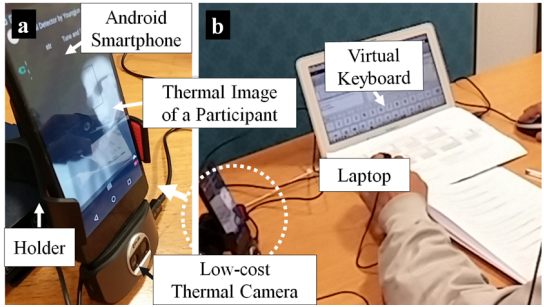

Figure. 5. Experimental setup: a smart phone with a low-cost thermal camera was installed and running on the desk while each person participated in the tasks. The participant was not asked to constrained their movemengs.

SD=10.22) were invited from the subject pool service of University College London. Each participant was given the information sheet and the informed consent form prior to data acquisition. The study took place in a quiet lab room with no distractions (and no temperature room control).

The flow for our experimental design is shown in Figure 4. After a 5-minute relaxation period (i.e., *Relax* in Figure 4), each participant was asked to go through the two types of cognitively demanding tasks (i.e., *Stroop, Math*). Before and after each demanding task (i.e., during *Q*), all participants were instructed to rate the subjective feeling based on a continuous 10-cm Visual Analogue Scale (VAS). VAS is a standard approach used to avoid the non-parametric property of the Likert scale [33]. In this study, the main question was: *"Did you feel mentally stressed?" (ranging from 0cm, not at all, to 10cm, very much)*. The easy and difficult sessions of each task type (i.e., *Stroop* 1,2 and *Math* 1,2) were counter-balanced.

The recording setup is shown in Figure 5. During the tests, each participant answered the task questions using a mouse on a laptop and was thermal-video-recorded using a low-cost thermal camera (FLIR One) connected to an Android smartphone. In the Stroop test condition, participants had to click-select the right answer among different colour options while pronouncing the colour aloud. Every question was shown for 1.5 seconds as in [32]. In the Math test condition, each participant typed an answer using a mouse on a virtual keypad (i.e., GUI in the screen). Each Math question was shown for 7.5 seconds which was set based on a pilot study.

The whole sessions took 63 minutes - 72 minutes per participant. The collected thermal videos were segmented according to the task and difficulty level they are related to. Using the techniques explained in Section 3.1 and 3.2, each video segment was transformed into an augmented set of spectrograms with the associated self-reported stress level (i.e., the scores reported at the end of each session). This process produced 3936 augmented input images of size 120 x 120 (i.e., an average of 492 (SD=91.49) for each 8 participant) which are publicly available in *http://youngjuncho.com/datasets/*.

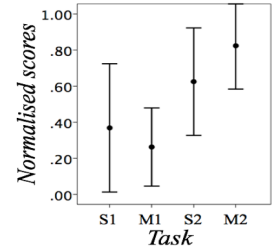

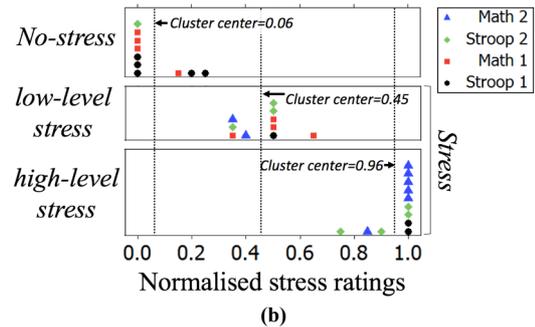

Figure. 6. Normalised subjective VAS scores: (a) distribution of the self report per participant (95% confidence interval), (b) clustered results using *k*-means. Colours and shapes of each point indicate the task type and difficulty levels. (Easy: Stroop 1(S1) and Math 1(M1); Difficult: Stroop 2(S2), Math 2(M2)).

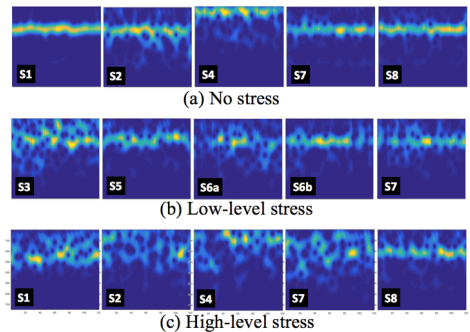

Figure. 7. Examples of respiration spectrograms clustered according to the participants' clustered VAS scores: (a) No-stress, (b) Low-level stress, (c) High-level stress.

## 5. Ground truth and stress classification

The classifier attempts to classify stress into one of three stress classes: *none*, *low* and *high*. We developed this classifier in two stages. First, we normalised participants' stress ratings to reflect inter-person variability in ratings (e.g., [34]). For each participant, the normalisation was based on feature scaling using his/her minimum and maximum scores. Second, we used k-mean algorithm to cluster the normalised scores into the three stress classes. We used *k*-means because it has been shown to be effective in dealing with self-reported ratings (e.g., [35],

[36]). The clusters were then used to label the breathing spectrogram dataset, i.e., each spectrogram was labelled using the cluster the person's self-report belongs to. Figure 6 summarises the results. As intended, the distribution of self-reports (Figure 6a) shows different levels of perceived stress across each task session. Figure 6b shows the clustering of the normalised subjective ratings into the stress classes. Here we should remark that we used the data collected from the four task sessions (i.e., S1: easy Stroop, M1: easy Math, S2: difficult Stroop, M2: difficult Math) but not the ones from the initial relax period (widely used as baseline). There are rationales for this: First, the tasks effectively induced the different stress levels from no stress to the high level of stress given that there were significant differences between each pair of clustered groups (all t-test comparisons had p<0.001; *none-low:* t(18)=8.939, *low-high:*t(20)=13.390, *none-high:* t(20)=23.651). On the other hand, there were no significant differences between subjective ratings collected from the initial relax period and those clustered into *no-stress* group (t(16)=1.211, p=0.244), suggesting the no-stress ratings to be a good baseline. Second, the recognition of non-stress in the relax periods would have been easier because of the lack of verbalisation and movement (e.g., mouse clicking) which all could affect the breathing patterns (independently of stress levels) [11]. Indeed, the tasks including the easy and difficult sessions had been purposely designed to verify the robustness of the approach to work in verbal contexts and unconstrained sedentary movement settings and the use of the relax periods as baseline would have reduced the validity of the evaluation.

Examples of the labelled data are shown in Figure 7. As discussed in the literature (e.g., [34]), physiological signals can vary strongly between people (e.g., S4 shows a particularly different spectrogram from the others, see Figure 7a). Although the breathing dynamics (i.e., variation patterns) are different between each individual, the higher the levels of stress a person felt, the more irregular the patterns appears to be. Nonetheless, it is not ignorable that the patterns labelled as high and low-level stress share some similarities. For this reason, we also undertook a binary classification problem by combining the low-and high-level stress into a *stress* class to discriminate events of *stress* from *no-stress*.

For the classification of breathing patterns using the presented architecture (see Section 3.3), the number (*n*) of kernels (i.e., filters) of the first convolutional layer was set to 12 for the three-class classification (i.e., three levels of stress classification) and 9 for the binary classification. The number (*j*) of kernels of the second convolutional layer was set to 24 and 18, respectively. In a training process, a fixed learning rate of 0.5 and a batch size of 50 were used for 300 epochs. For the in-depth evaluation of the performance of the proposed *deep* model, the three single-layer neural networks were implemented as shallow learning methods. The first single-layer neural network (denoted as *NN1*) uses the resized image patches (i.e., 28x28 in the proposed model, see Figure 3) as input and includes one hidden layer of node size 30. The second shallow network (denoted as *NN2*) uses half-sized image patches (60x60) and a hidden layer of size 45. The last network (denoted as *NN3*) uses the original images (120x120) and a hidden layer of size 300. These shallow learning methods were run for both the binary and the three-class classification models. Finally, for the validation of all the networks, an 8-fold leave-one-*subject*-out (LOSO) cross-validation was used. At every fold, all data sequences from the participants except for one were used to train the networks, and the data from the left out participant was used for testing.

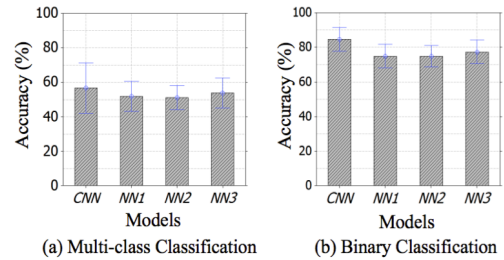

Figure. 8. Summary of classification accuracy for (a) three-class classification (i.e., none, low, high-level stress), (b) binary classification (i.e., no-stress, stress). CNN: 5 layers CNN using low resolution RVS images (28x28), NN1: single hidden layer using low resolution RVS images (28x28), NN2: single hidden layer using medium resolution RVS images (60x60), NN3: single hidden layer using high resolution RVS images (120x120).

|  | Predicted Class | | | | Predicted Class | | |
|--|No-stress|Low-level|High-level| |No-stress|Low-level|High-level|
|No-stress|731 (56.0%)|286 (21.9%)|288 (22.1%)|No|432 (33.1%)|463 (35.5%)|410 (31.4%)|
|Low-level stress|282 (23.4%)|465 (38.5%)|460 (38.1%)|Low|120 (10.0%)|598 (49.5%)|489 (40.5%)|
|High-level stress|0 (0.00%)|387 (27.2%)|1037 (72.8%)|High|182 (12.8%)|246 (17.3%)|996 (69.9%)|

(a) 5 layers CNN    (b) Single layer NN

Figure. 9. Results from the three-class classification problem (None, Low, High-level stress): (a) 5 layers CNN (Deep Learning), (b) single layer neural network (Shallow Learning). Confusion matrices (each block consists of the sum and average over the LOSO cross validation results, the colour represents the number of each prediction (i.e., black: highest, white: lowest)).

|  | Predicted Class | | | Predicted Class | |
|--|No-stress|Stress| |No-stress|Stress|
|No-stress|713 (54.6%)|592 (45.4%)|No-stress|467 (35.8%)|838 (64.2%)|
|Stress|128 (05.9%)|2503 (94.1%)|Stress|225 (08.6%)|2406 (91.4%)|

(a) 5 layers CNN    (b) Single layer NN

Figure. 10. Results from the binary classification problem (No-stress and stress): (a) 5 layers CNN, (b) single layer neural network. Confusion matrices (same color coding as in Figure 9).

## 6. Results

The dataset is composed of the augmented breathing spectrogram sequences of size 120 (i.e., frequency range) x 120 (window length) x 3936 (the number of two-dimensional spectrograms) and the stress labels (i.e., either with binary or three-class labels). Figure 8 summarises the accuracy results from the leave-one-subject-out cross validation (N=8) for both the three-class and the binary cases. The CNN results were compared with NN1, NN2, NN3 results to understand the contribution made by the deep learning method.

**Multi-levels stress recognition:** the highest accuracy (M=56.52%, SD=17.58%) was achieved with the deep CNN model, while M=51.89% (SD=10.36%), M=51.02% (SD= 8.33%), 53.65% (SD=10.39%) were the results from NN1, NN2, NN3, respectively. The average F1 scores across every participant were 61.33% (SD=19.71%), 57.64% (SD=13.90%), 57.57% (SD=11.80%), 59.84% (SD=14.23%) for the proposed model, NN1, NN2, NN3 respectively. When comparing the deep CNN and NN1 which deal with the same input images of size 28x28, we found that the deep CNN was capable of predicting both the *no-stress* and the *high-level-stress* better than NN1 while NN1 produced a better result in predicting the *low-level stress* as shown in Figure 9.

**Binary stress recognition:** the deep CNN produced a much higher accuracy (M=84.59%, SD=19.34%) than the shallow networks (NN1: M=74.89%, SD= 19.12%, NN2: M=74.81%, SD=17.58%, NN3: M=77.31%, SD=19.45%). The average F1 scores across every participant were 87.22% (SD=16.32%), 75.18% (SD=21.96%), 79.14% (SD=14.81%), 82.04% (SD=15.81%) for NN1, NN2, NN3 respectively. Compared to NN1, the deep CNN performed significantly better in predicting both the *no-stress* (713 corrected predictions from CNN and 467 from NN1) *and stress* states (2503 from CNN and 2406 from NN1) as shown in Figure 10.

## 7. Discussions

The main aim of this paper was to contribute the design of a novel low-cost non-contact thermography based stress recognition system by going deeper into breathing dynamic patterns. With this goal, we also contribute a new input signature: respiration variation spectrogram condensing breathing dynamics, an automatic feature learning method supported by a deep learning framework, and a new thermal image dataset labelled with the stress levels from our structured multi-level stress induction tasks.

The accuracy results of our study, show that the proposed deep learning-based stress recognition model produces better performances (84.59% (binary), 56.52% (multi-class)) than the best results from the shallow learning method (single-layer NN: 77.31% (binary), 53.65% (multi-class)). However, those accuracy results were not significantly different. This may be related to the relatively small number of participants although this number was set reflecting [32]. Regardless of this differentiation of deep and shallow networks made in this evaluation, this is the first study aimed at learning breathing patterns to automatically recognise a person's mental stress level without hand-engineered feature extractors which require in-depth domain knowledge and highly matured skills. In particular, our proposed respiration variability spectrogram inputs (RVSs) are able to capture the dynamic breathing information from a very short-lasting (e.g., a sudden drop and increase) to long-lasting respiration changes. The proposed convolutional kernels based model can support the feature learning of the inputs, and in turn, contribute to the accuracy of a unimodal model.

Compared with the literature suggesting low discriminative power of breathing in discriminating between stress levels [10], the proposed model shows that respiration signals alone can lead to above chance levels in 3-classes recognition (i.e., high, low, no stress). In addition, in the case of binary recognition our performances are similar to the one reported in [10] with the difference that in [10] multiple physiological channels were used rather than just breathing. We suggest that our high performances are partially due to the use of breathing variability rather than just breathing rate. We hence expect that the combination of breathing variability with other signals may lead to further improvement in the recognition rate.

Further improvements could be obtained through a more refined ground truth (e.g., continuous over each session), improved normalisation of the subjective ratings and by considering inter-person variability (e.g., Figure 7). Finally, despite the fact that the evaluation tasks did not require large movement, people were free to move as needed (e.g., inputting their responses, or displaying stress related expressions e.g. turning or bending the head to escape the source of stress and frustration [37]). We expect, that even larger movements would be acceptable as the ROI tracking was tested in [15] in highly mobile contexts and high thermal environmental changes.

## 8. Conclusion

DeepBreath is a novel stress recognition system based on deep learning and low-cost thermal imaging. Our main contributions – a new bi-dimensional signature condensing complex breathing dynamics, a new data augmentation method-based deep feature learning and the exploration of the low-cost thermal imaging for real-world ubiquitous affect recognition - were thoroughly evaluated through a multi-level, multi-task stress induction study. To the best of our knowledge, this is the first use of a deep learning-applied approach on breathing dynamic patterns. The results show high performances in automatic stress detection under unconstrained settings (i.e., verbal expressions and movement).

**Acknowledgments**

Youngjun Cho was supported by University College London Overseas Research Scholarship (UCL-ORS) awarded to top quality international postgraduate students. This work was partially supported by H2020-Human Project (ID: 723737 - http://www.humanmanufacturing.eu/).

## References

[1] B. S. McEwen, "Physiology and Neurobiology of Stress and Adaptation: Central Role of the Brain", *Physiol. Rev.*, vol. 87, no. 3, pp. 873–904, Jul. 2007.

[2] J. M. Nash and R. W. Thebarge, "Understanding Psychological Stress, Its Biological Processes, and Impact on Primary Headache", *Headache J. Head Face Pain*, vol. 46, no. 9, pp. 1377–1386, Oct. 2006.

[3] A. F. T. Arnsten, "Stress signalling pathways that impair prefrontal cortex structure and function", *Nat. Rev. Neurosci.*, vol. 10, no. 6, pp. 410–422, Jun. 2009.

[4] P. J. Pretty, J. Peacock, M. Sellens, and M. Griffin, "The mental and physical health outcomes of green exercise", *Int. J. Environ. Health Res.*, vol. 15, no. 5, pp. 319–337, Oct. 2005.

[5] H. Lu et al., "StressSense: Detecting Stress in Unconstrained Acoustic Environments Using Smartphones", in *Proceedings of the 2012 ACM Conference on Ubiquitous Computing*, 2012, pp. 351–360.

[6] J. A. Healey and R. W. Picard, "Detecting stress during real-world driving tasks using physiological sensors", *IEEE Trans. Intell. Transp. Syst.*, vol. 6, no. 2, pp. 156–166, Jun. 2005.

[7] R. S. Lazarus, "From psychological stress to the emotions: A history of changing outlooks", *Annu. Rev. Psychol.*, vol. 44, p. 1, 1993.

[8] J. F. Brosschot, W. Gerin, and J. F. Thayer, "The perseverative cognition hypothesis: A review of worry, prolonged stress-related physiological activation, and health", *J. Psychosom. Res.*, vol. 60, no. 2, pp. 113–124, Feb. 2006.

[9] P. Grossman, "Respiration, Stress, and Cardiovascular Function', *Psychophysiology*, vol. 20, no. 3, pp. 284–300, May 1983.

[10] D. J. McDuff, J. Hernandez, S. Gontarek, and R. W. Picard, "COGCAM: Contact-free Measurement of Cognitive Stress During Computer Tasks with a Digital Camera", in *Proceedings of the 2016 CHI Conference on Human Factors in Computing Systems*, 2016, pp. 4000–4004.

[11] J.-H. Hong, J. Ramos, and A. K. Dey, "Understanding Physiological Responses to Stressors During Physical Activity", in *Proceedings of the 2012 ACM Conference on Ubiquitous Computing*, 2012, pp. 270–279.

[12] Y. LeCun, Y. Bengio, and G. Hinton, "Deep learning", *Nature*, vol. 521, no. 7553, pp. 436–444, May 2015.

[13] R. Murthy and I. Pavlidis, "Noncontact measurement of breathing function", *IEEE Eng. Med. Biol. Mag.*, vol. 25, no. 3, pp. 57–67, May 2006.

[14] C. B. Pereira, X. Yu, M. Czaplik, R. Rossaint, V. Blazek, and S. Leonhardt, "Remote monitoring of breathing dynamics using infrared thermography", *Biomed. Opt. Express*, vol. 6, no. 11, pp. 4378–4394, Nov. 2015.

[15] Y. Cho, S. J. Julier, N. Marquardt, and N. Bianchi-Berthouze, "Robust tracking of respiratory rate in high-dynamic range scenes using mobile thermal imaging", *Biomed. Opt. Express,* 2017.

[16] Y. Masaoka and I. Homma, "Anxiety and respiratory patterns: their relationship during mental stress and physical load", *Int. J. Psychophysiol.*, vol. 27, no. 2, pp. 153–159, Sep. 1997.

[17] H. Genno et al., "Using facial skin temperature to objectively evaluate sensations", *Int. J. Ind. Ergon.*, vol. 19, no. 2, pp. 161–171, Feb. 1997.

[18] V. Engert, A. Merla, J. A. Grant, D. Cardone, A. Tusche, and T. Singer, "Exploring the Use of Thermal Infrared Imaging in Human Stress Research", *PLOS ONE*, vol. 9, no. 3, p. e90782, Mar. 2014.

[19] C. Puri, L. Olson, I. Pavlidis, J. Levine, and J. Starren, "StressCam: Non-contact Measurement of Users' Emotional States Through Thermal Imaging", in *CHI '05 EA,* 2005, pp. 1725–1728.

[20] B. R. Nhan and T. Chau, "Classifying Affective States Using Thermal Infrared Imaging of the Human Face", *IEEE Trans. Biomed. Eng.*, vol. 57, no. 4, pp. 979–987, Apr. 2010.

[21] M. M. Khan, R. Ward, and M. Ingleby, "Toward Use of Facial Thermal Features in Dynamic Assessment of Affect and Arousal Level", *IEEE Tr.. Affect. Comp..*, vol. PP, no. 99, pp. 1–1, 2016.

[22] A. Krizhevsky, I. Sutskever, and G. E. Hinton, "ImageNet Classification with Deep Convolutional Neural Networks", in *Advances in Neural Information Processing Systems 25*, 2012, pp. 1097–1105.

[23] C. Szegedy et al., "Going Deeper With Convolutions", in *the Proceedings of the IEEE Conference on Computer Vision and Pattern Recognition*, 2015, pp. 1–9.

[24] P. Liu, S. Han, Z. Meng, and Y. Tong, "Facial Expression Recognition via a Boosted Deep Belief Network", in *the Proceedings of the IEEE Conference on Computer Vision and Pattern Recognition*, 2014, pp. 1805–1812.

[25] N. D. Lane, P. Georgiev, and L. Qendro, "DeepEar: Robust Smartphone Audio Sensing in Unconstrained Acoustic Environments Using Deep Learning", in *ACM International Joint Conference on Pervasive and Ubiquitous Computing*, 2015, pp. 283–294.

[26] M. Lin, Q. Chen, and S. Yan, "Network In Network", *arXiv preprint arXiv: 1312.4400*, 2013.

[27] Y. LeCun and Y. Bengio, "Convolutional Networks for Images, Speech, and Time Series", in *Handbook of Brain Theory and Neural Networks*, M. A. Arbib, Ed. MIT Press, 1995, p. 3361.

[28] G. Trigeorgis et al., "Adieu features? End-to-end speech emotion recognition using a deep convolutional recurrent network", in *2016 IEEE Inter. Conf. on Acoustics, Speech and Signal Processing (ICASSP)*, 2016, pp. 5200–5204.

[29] S. Ghosh, E. Laksana, S. Scherer, and L. P. Morency, "A multi-label convolutional neural network approach to cross-domain action unit detection", in *Inter. Conf. on Affective Computing and Intelligent Interaction (ACII)*, 2015, pp. 609–615.

[30] Z. Yu and C. Zhang, "Image Based Static Facial Expression Recognition with Multiple Deep Network Learning", in *ACM on Inter. Conf. on Multimodal Interaction*, 2015, pp. 435–442.

[31] J. R. Stroop, "Studies of interference in serial verbal reactions", *J. Exp. Psychol.*, vol. 18, no. 6, pp. 643–662, 1935.

[32] T. Åkerstedt et al., "Comparison of urinary and plasma catecholamine responses to mental stress", *Acta Physiol. Scand.*, vol. 117, no. 1, pp. 19–26, Jan. 1983.

[33] P. E. Bijur, W. Silver, and E. J. Gallagher, "Reliability of the Visual Analog Scale for Measurement of Acute Pain", *Acad. Emerg. Med.*, vol. 8, no. 12, pp. 1153–1157, Dec. 2001.

[34] J. Hernandez, R. R. Morris, and R. W. Picard, "Call Center Stress Recognition with Person-Specific Models", in *Affective Computing and Intelligent Interaction*, 2011, pp. 125–134.

[35] C. Salmivalli, A. Kaukiainen, L. Kaistaniemi, and K. M. J. Lagerspetz, "Self-Evaluated Self-Esteem, Peer-Evaluated Self-Esteem, and Defensive Egotism as Predictors of Adolescents' Participation in Bullying Situations", *Pers. Soc. Psychol. Bull.*, vol. 25, no. 10, pp. 1268–1278, Oct. 1999.

[36] K. Kjeldstadli et al., "Life satisfaction and resilience in medical school – a six-year longitudinal, nationwide and comparative study", *BMC Med. Educ.*, vol. 6, p. 48, Sep. 2006.

[37] A. Kleinsmith and N. Bianchi-Berthouze, "Affective Body Expression Perception and Recognition: A Survey", *IEEE Trans. Affect. Comput.*, vol. 4, no. 1, pp. 15–33, Jan. 2013.